\documentclass[runningheads]{llncs}

\usepackage[T1]{fontenc}
\usepackage[utf8]{inputenc} 
\usepackage{amsmath}
\usepackage{booktabs}
\usepackage{amssymb}
\usepackage{xcolor}
\usepackage[numbers]{natbib}
\usepackage{array} 
\usepackage{hyperref}
\usepackage{graphicx}
\usepackage{ragged2e}

\begin{document}
\title{Performance Evaluation of Deep Learning and Transformer Models Using Multimodal Data for Breast Cancer Classification}

\author{Sadam Hussain\inst{1}\orcidID{0000-0002-3453-0785} \and
Mansoor Ali\inst{1}\orcidID{0000-0002-4669-7267} \and
Usman Naseem\inst{2}\orcidID{0000-0003-0191-7171}\and
Beatriz Alejandra Bosques Palomo \inst{1} \and Mario Alexis Monsivais Molina \inst{1} \and Jorge Alberto 
Garza Abdala \inst{1} \and
Daly Betzabeth Avendano Avalos\inst{4}\and
Servando Cardona-Huerta\inst{4} \and T. Aaron Gulliver \inst{3} \and
Jose Gerardo Tamez Pena\inst{4}\orcidID{0000-0003-1361-5162}}
%
%
\institute{Tecnologico de Monterrey, School of Sciences and Engineering, Mexico.\\ \email{a01753094@tec.mx} \and
School of Computing, Macquarie University, Australia. \and Department of Electrical and Computer Engineering, University of Victoria, Canada. \and
School of Medical and Health Sciences, Tecnologico de Monterrey, Mexico.}

\maketitle              
\begin{abstract}

Rising breast cancer (BC) occurrence and mortality are major global concerns for women. Deep learning (DL) has demonstrated superior diagnostic performance in BC classification compared to human expert readers. However, the predominant use of unimodal (digital mammography) features may limit the current performance of diagnostic models. To address this, we collected a novel multimodal dataset comprising both imaging and textual data. This study proposes a multimodal DL architecture for BC classification, utilizing images (mammograms; four views) and textual data (radiological reports) from our new in-house dataset. Various augmentation techniques were applied to enhance the training data size for both 
imaging and textual data. We explored the performance of eleven SOTA DL architectures (VGG16, VGG19, ResNet34, ResNet50, MobileNet-v3, EffNet-b0, EffNet-b1, 
EffNet-b2, EffNet-b3, EffNet-b7, and Vision Transformer (ViT)) as imaging feature extractors. For textual feature extraction, we utilized either artificial neural networks (ANNs) or long short-term memory (LSTM) networks. The combined imaging and textual features were then inputted into an ANN classifier for BC classification, using the late fusion technique. We evaluated different feature extractor and classifier arrangements. The VGG19 and ANN combinations achieved the highest accuracy of 0.951. For precision, the VGG19 and ANN combination again surpassed other CNN and LSTM, ANN based architectures by achieving a score of 0.95. The best sensitivity score of 0.903 was achieved by the VGG16+LSTM. The highest F1 score of 0.931 was achieved by VGG19+LSTM. Only the VGG16+LSTM achieved the best area under the curve (AUC) of 0.937, with VGG16+LSTM closely following with a 0.929 AUC score.

\end{abstract}

\section{Introduction}
BC is the most prevalent disease among women \cite{sung2021global}. Anticipated figures for 2023 suggest that the United States will likely witness 1,958,310 new cancer diagnoses and 609,820 cancer-related fatalities \cite{Hansebout2009HowTU}. Digital mammography and other imaging modalities have long been used for BC detection \cite{hussain2023deep}. However, due to the large volume of data, radiologists often struggle to process it in a timely manner. To assist them, various Computer-Aided Diagnosis (CAD) systems have been developed \cite{yassin2018machine, calisto2022breastscreening}.

Traditionally, many CAD systems are tailored to process single-modality data, such as images, text, or audio. However, recent studies indicate that using multimodal data as input to SOTA DL models can yield promising results in the medical field, particularly in BC diagnosis and prognosis \cite{huang2020multimodal, tomczak2015review, gao2013integrative, sun2018multimodal}. Multimodal data is crucial for BC diagnosis and prognosis, as it provides comprehensive information about various aspects such as radiomics characteristics, clinical features, and imaging features of the disease. Integrating different types of data allows DL methods to achieve better performance in BC classification, prognosis, and survival prediction than using only a single modality \cite{huang2020multimodal}. 

Multimodal BC classification has received increasing attention of researchers. Most approaches use imaging data along with textual reports for BC diagnosis \cite{akselrod2019predicting, holste2021end} or some studies such as \cite{holste2021end} used clinical factors. Some studies employed multidimensional data for predicting both short-term and long-term BC risk \cite{sun2018multimodal}. 

In this study, we provide a comparative analysis of various SOTA DL models with ViT \cite{dosovitskiy2020image}. We also collected a multimodal dataset comprising digital mammography (four views; L-CC, L-MLO, R-CC and R-MLO) and radiological reports. We propose a DL and transformer-based pipeline with different backbone networks to extract features from the multimodal data, applying a late fusion technique to combine imaging and textual features before training a classifier for BC classification. We also evaluated the performance of the transformer-based architecture, ViT, which has shown promising outcomes on text and imaging data \cite{fields2023vision}.

\begin{figure*}
\center
  \includegraphics[width=1.0\textwidth]{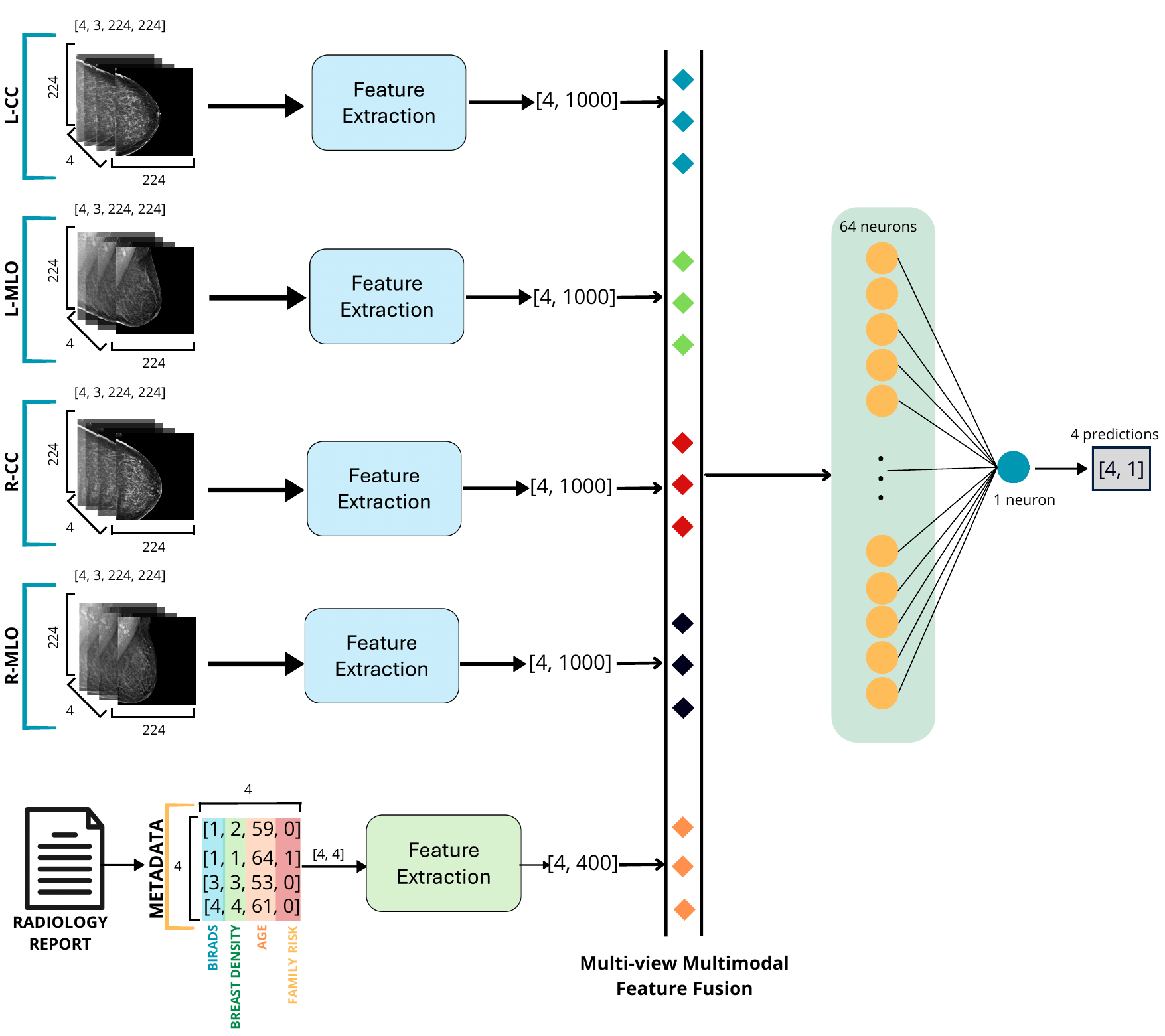}
  \caption{Proposed Multi-Modal Model for Processing Multi-Dimensional Data}
  \label{fig1}
\end{figure*}

The proposed model uses CNNs and ViT for feature extraction from digital mammography and either ANN or LSTM for feature extraction from radiological reports. The fused features are then assigned to an ANN classifier for the final classification of benign or malignant cancer. We observed that models combining LSTM for textual feature extraction generally performed better than those using simple ANN. Furthermore, it has been observed that VGG16 and VGG19 models outperformed other SOTA DL architectures and ViT across all the evaluated metrics The proposed multimodal pipeline is illustrated in Fig. \ref{fig1}.

\section{Proposed Work}
In this work, we evaluate the performance of various DL and transformer based architectures on a newly collected multimodal dataset for BC diagnosis. 

The key contributions of this work are highlighted as follows:

i. A new in-house dataset comprising multi-view mammograms and radiological reports has been collected and preprocessed.

ii. A DL and transformer-based multimodal pipeline for BC classification is employed on a new in-house dataset. In this pipeline, SOTA DL models, such as VGG16 \cite{simonyan2014very}, VGG19 \cite{simonyan2014very}, ResNet34 \cite{he2016deep}, ResNet50 \cite{he2016deep},  MobileNet-v3 \cite{howard2019searching}, EffNet-b0 \cite{tan2019efficientnet}, EffNet-b1 \cite{tan2019efficientnet}, EffNet-b2 \cite{tan2019efficientnet}, EffNet-b3 \cite{tan2019efficientnet}, EffNet-b7 \cite{tan2019efficientnet}, and a transformer-based model, ViT \cite{dosovitskiy2021image}, are used to extract imaging features, with either LSTM or ANN employed to extract textual features.

\section{Methodology}
This section highlights the methodology employed in this study, encompassing data collection, preprocessing, data augmentation, as well as the training and testing of data. It also details the multimodal architecture utilized, implementation specifics, classification, and evaluation of the models.

\subsection{Data}
The dataset used in this investigation was provided by the breast radiology department at TecSalud Hospital in Monterrey, Mexico. In addition to digital mammograms, it includes digital breast tomosynthesis (DBT), digital mammography, and ultrasound (US) based radiological reports. Anonymized reports and mammograms were used to ensure anonymity. The reports, which were gathered between January and December 2014 and were initially written in Spanish, were translated into English using Google Translate. Bilingual radiologists then confirmed the translation. Two seasoned radiologists wrote and evaluated each report. One radiologist reviewed the translated reports. In cases where the first radiologist disagreed with the translation, the second radiologist revised and finalized the translation. There were a thousand examples in the dataset at first. Four views (LCC, LMLO, RCC, and RMLO) and related textual data (BIRADS categories, breast density scores, patient ages, family cancer histories, and lesion laterality) make up each patient's data set. Lesion laterality reveals if a patient has had cancer in one or both breasts. Three values are used to characterize it: 0 for left breast, 1 for right breast, and 2 for cancer in both breasts. Laterality was only performed to determine whether the patient had malignancy. We used 770 cases with a total of 3080 mammograms and related metadata after preprocessing, which comprised deleting reports with missing values, duplicates, and BIRADS categories 0 (inconclusive) and 6 (biopsy-proven). There are exactly equal numbers of positive (385) and negative (385) cases in the dataset. The final model building and evaluation involved women in all cases, with an average patient age of 53.

\subsection{Data Preprocessing}
The original dataset, in Spanish, consisted of 7,904 samples. After translation, we removed redundant data, missing values, and cases with BIRADS scores of 0 or 6. Following preprocessing, we collected 5,046 samples. Each patient entry included a brief clinical indication of the study, personal risk history, imaging description, findings, and diagnostic impression. From these radiological reports, we extracted BIRADS categories (1-5), breast density scores, patients' ages, and family history of cancer, which were ultimately used for model building.

\subsection{Data Augmentation}
The augmentation techniques applied to the mammogram dataset were carefully selected to introduce diversity and robustness into the training data. Horizontal and vertical flipping were utilized to mirror the images, exposing the model to different orientations. Furthermore, the implementation of random resized crops, targeting an output size of 224x224 pixels, aimed to instill scale and translation invariance during model training. Dynamic transformations, such as shifts, scales, and rotations, were also incorporated, allowing for rotations up to 90 degrees and scale adjustments within the range of 0.8 to 1.2. This diverse set of augmentations serves the crucial purpose of expanding the effective size of the training dataset. By exposing the model to varied perspectives, the augmentation strategy becomes instrumental in preventing overfitting and enhancing the model's adaptability to diverse input scenarios, ultimately contributing to better performance.

\subsection{Multimodal Deep Architecture}

We introduced a multimodal architecture to classify BC instances as positive or negative. The proposed architecture consists of independent DL and transformer-based architectures (pre-trained on ImageNet "IMAGENET1K\_V1" weights provided by the torchvision library) as a backbone for extracting imaging features from mammograms. The DL architectures used in this study are variants of VGG, ResNet, EffNet and MobileNetv3. We chose these techniques because of their SOTA performance on natural and medical image classification tasks \citep{tan2019efficientnet, xu2023resnet, ikechukwu2021resnet, dosovitskiy2021image, qian2021mobilenetv3}. For extracting textual features that were extracted from radiological reports as tabular data we used simple ANN or long short-term memory (LSTM). These simple architectures perform better on tabular data. Afterwards, the feature vectors extracted from both modalities were fused before the final classification task. This vector is then passed through a linear classification layer for the prediction of malignant or benign cancer. 

Let \(\mathcal{I}\) represents mammogram input, where there are four input images one for each view of the mammogram, therefore, $\textit{i}_1$, $\textit{i}_2$, $\textit{i}_3$ and $\textit{i}_4$, represents each view of the mammogram respectively. Features extracted from the four mammogram views are represented as $\textit{f}_1$, $\textit{f}_2$, $\textit{f}_3$ and $\textit{f}_4$. Let \(\mathcal{T}\) represents input tabular data extracted from radiological reports. Features extracted from tabular data are shown as $\textit{f}_t$. Therefore, the concatenated feature vector represented as \(\mathcal{F}_c\) = concat($\textit{f}_{1}$, $\textit{f}_{2}$, $\textit{f}_{3}$, $\textit{f}_{4}$, $\textit{f}_t$). The \(\mathcal{F}_c\) is then fed to a linear layer for BC classification.    

The transformer architecture is an attention-based encoder-decoder model. The encoder processes the input, and the decoder generates the output. Importantly, the transformer architecture does not depend on recurrence or convolutions to produce an output. 
The architecture employs stacked self-attention and point-wise, fully connected layers for both the encoder and decoder, as detailed in Vaswani et al., 2017 \cite{vaswani2017attention}.

The self-attention mechanism is one of the most important parts of the transformer architecture. It makes the model to consider the relationships between all positions in the input simultaneously with:

\begin{equation}
    MultiHead(Q, K, V) = Concat(head_1, ..., head_h)W^O
\end{equation}
  \label{eqn1}
\begin{equation}
    head_i = Attention(QW_i^Q, KW_i^K, VW_i^V),
\end{equation}
  \label{eqn2}
where Q is the query, K denotes to the key, and V is the value \cite{vaswani2017attention}.

The ViT is an extension of the transformer architecture, utilizing only the encoder part. The input image is divided into fixed-size, non-overlapping patches. Each patch is linearly embedded to form vectors, with an added "classification token." Thus, the transformer's input is the classification token combined with the patch embeddings. The transformer's output then serves as the input for a classification head.

\subsection{Implementation Details}
For training and testing all models, we utilized an NVIDIA RTX A4000 GPU. This GPU is equipped with 16 GB of memory and supports PCI Express Gen 4. Additionally, it features third-generation tensor CUDA cores. Our network architecture incorporated ResNet34, ResNet50, VGG16, VGG19, MobileNet\_v3, ViT, and EffNet\_b0 as backbones for image encoding, and either ANN or LSTM for text encoding. The model weights were initialized from scratch rather than using pretrained models. We employed Adam as the optimizer to minimize loss, with a learning rate set at 0.0005. The batch size was fixed at 8, and the models were trained over 100 epochs. The Rectified Linear Unit (ReLU) was used as the activation function, with a dropout rate of 0.2. The dataset was split into training and validation sets at a ratio of 80\% to 20\%, respectively. This data split was consistent across all experiments. Model performance was evaluated using the validation set.

\section{Results}
In our comprehensive evaluation, we used a variety of performance metrics to assess the effectiveness of our multimodal models, ensuring a thorough understanding of their classification capabilities. These metrics included Accuracy, Precision, Sensitivity, F1 Score, and AUC. Our results demonstrated significant achievements across different metrics, highlighting the distinct strengths of various models.

The highest accuracy, at 0.951, was achieved by the VGG19+ANN architecture, indicating the superior ability to correctly classify instances within the dataset. This was closely followed by the joint models of VGG19+LSTM, scoring 0.919. For precision, the VGG19 paired with ANN achieved the highest score of 0.95, surpassing other combined model configurations, followed by the ViT and LSTM combination scoring 0.932. This highlights VGG19's proficiency in minimizing false positives and ensuring precise classifications. In terms of sensitivity, the VGG16 and LSTM combination recorded the highest rate at 0.903, followed by VGG19+LSTM, at 0.900. This indicates their effectiveness in correctly identifying a significant proportion of true positive cases, crucial in applications where sensitivity is key. The highest F1 Score, reflecting a balance between precision and recall, was 0.931 using the VGG19+LSTM, with the VGG19+ANN close behind at 0.922. The AUC analysis showed the VGG16+LSTM model leading with a score of 0.937, demonstrating its ability to differentiate between positive and negative cases effectively, followed by the VGG16 and ANN combination at 0.929. The results of combining SOTA DL architectures with LSTM for BC classification are presented in Table \ref{tab1}, and those with ANN in Table \ref{tab2}.

The least performing model was the ResNet34 combined with ANN, recording an accuracy of 0.692. The MobileNet\_v3 combined with ANN was the least accurate in the LSTM category, at 0.813. The EffNet\_b3 paired with ANN had the lowest precision of 0.711, while ResNet50 with LSTM scored the lowest at 0.747. In terms of sensitivity, the lowest score, 0.643, was recorded by the MobileNet\_v3 and ANN combination, followed by the EffNet\_b3 combined with ANN scoring 0.653. The EffNet\_b3 and ANN combination had the lowest F1 Score at 0.652, while the lowest in the LSTM category was 0.723 for EffNet\_b0, EffNet\_b1, EffNet\_b2. Lastly, the lowest AUC score of 0.678 was achieved by the MobileNet\_v3 and ANN combination, and 0.679 by the ViT and LSTM combination.

\begin{table*}[t!]
\caption{Overview of the Results of Different DL Models for BC Classification}
\begin{center}
\resizebox{\textwidth}{!}{%
\begin{tabular}{l|l|l|l|l|l|l}
\hline
\textbf{Feature Extractor} & \textbf{Classifier} & \textbf{ACC} & \textbf{Precision} & \textbf{Sensitivity} & \textbf{F1 Score} & \textbf{AUC} \\ \hline
VGG16+LSTM & ANN & 0.882 & 0.902 & \textbf{0.903} & 0.902 & \textbf{0.937} \\ \hline
VGG19+LSTM & ANN & \textbf{0.919} & 0.923 & 0.900 & \textbf{0.931} & 0.907 \\ \hline
ResNet34+LSTM & ANN & 0.910 & 0.782 & 0.899 & 0.753 & 0.739 \\ \hline
ResNet50+LSTM & ANN & 0.889 & 0.747 & 0.899 & 0.753 & 0.798 \\ \hline
MobileNet\_v3+LSTM & ANN & 0.813 & 0.835 & 0.753 & 0.903 & 0.679 \\ \hline
EffNet\_b0+LSTM & ANN & 0.834 & 0.769 & 0.892 & 0.723 & 0.757 \\ \hline
EffNet\_b1+LSTM & ANN & 0.834 & 0.772 & 0.895 & 0.723 & 0.759 \\ \hline
EffNet\_b2+LSTM & ANN & 0.842 & 0.893 & 0.876 & 0.723 & 0.807 \\ \hline
EffNet\_b3+LSTM & ANN & 0.827 & 0.758 & 0.903 & 0.736 & 0.719 \\ \hline
EffNet\_b7+LSTM & ANN & 0.825 & 0.884 & 0.872 & 0.738 & 0.685 \\ \hline
ViT+LSTM & ANN & 0.893 & \textbf{0.932} & 0.863 & 0.871 & 0.925 \\ \hline
\end{tabular}
}
\label{tab1}
\end{center}
\end{table*}

\begin{table*}[t!]
\caption{Overview of the Results of Different DL Models for BC Classification}
\begin{center}
\resizebox{\textwidth}{!}{%
\begin{tabular}{l|l|l|l|l|l|l}
\hline
\textbf{Feature Extractor} & \textbf{Classifier} & \textbf{ACC} & \textbf{Precision} & \textbf{Sensitivity} & \textbf{F1 Score} & \textbf{AUC} \\ \hline
VGG16+ANN & ANN & 0.900 & 0.912 & \textbf{0.893} & 0.897 & \textbf{0.929} \\ \hline
VGG19+ANN & ANN & \textbf{0.951} & \textbf{0.950} & 0.884 & \textbf{0.922} & 0.915 \\ \hline
ResNet34+ANN & ANN & 0.692 & 0.731 & 0.716 & 0.692 & 0.748 \\ \hline
ResNet50+ANN & ANN & 0.721 & 0.752 & 0.705 & 0.681 & 0.723 \\ \hline
MobileNet\_v3+ANN & ANN & 0.754 & 0.858 & 0.643 & 0.668 & 0.678 \\ \hline
EffNet\_b0+ANN & ANN & 0.781 & 0.714 & 0.658 & 0.668 & 0.709 \\ \hline
EffNet\_b1+ANN & ANN & 0.781 & 0.714 & 0.658 & 0.668 & 0.703 \\ \hline
EffNet\_b2+ANN & ANN & 0.782 & 0.887 & 0.825 & 0.851 & 0.828 \\ \hline
EffNet\_b3+ANN & ANN & 0.781 & 0.711 & 0.653 & 0.652 & 0.719 \\ \hline
EffNet\_b7+ANN & ANN & 0.781 & 0.812 & 0.658 & 0.687 & 0.689 \\ \hline
ViT+ANN & ANN & 0.871 & 0.903 & 0.819 & 0.848 & 0.844 \\ \hline
\end{tabular}
}
\label{tab2}
\end{center}
\end{table*}

\section{Discussion}
The data in healthcare is inherently multimodal\cite{acosta2022multimodal}. It includes scans in the form of images, doctors' analyses as notes or radiological reports, and audio data, such as ECG/EEG, which provide insights for diagnosis, prognosis, and treatment decisions\cite{pei2023review}. However, most healthcare studies focus on a single modality of data (image, text, or audio) due to the scarcity of multimodal data, often yielding less than optimal results. Since the last decade, researchers have been combining different data modalities to achieve better outcomes. Current methods, though, still fall short in terms of generalizability and accuracy comparable to that of physicians\cite{heiliger2022beyond}.

In this work, we extracted features from images and textual data, combining them for the final classification of benign and malignant cancer. We employed various SOTA DL and transformer-based architectures for feature extraction from mammograms and radiological reports. These included VGG16, VGG19, ResNet34, ResNet50, MobileNet\_v3, EffNet\_b0, EffNet\_b1, EffNet\_b2, EffNet\_b3, EffNet\_b7, and ViT for mammogram feature extraction. For textual feature extraction, we used ANN or LSTM techniques. We then merged these features using a late fusion strategy and employed an ANN architecture for classification. ViT, in combination with LSTM or ANN, couldn't achieve better or comparable results to other SOTA DL models in terms of accuracy, precision, sensitivity, F1 score and AUC. However, ViT+LSTM ranked second highest in the AUC curve with a score of 0.925, just behind VGG16+LSTM with the highest score of 0.937.

We observed that using LSTM for textual feature extraction with any DL architecture as a backbone for imaging feature extraction can slightly enhance the overall model's performance in terms of sensitivity, F1 score and AUC, compared to basic ANN. However, ANN combinations outperformed LSTM in metrics like accuracy and precision. Both ANN and LSTM effectively utilized the rich contextual information in textual data, improving the model's capability to discern complex patterns and relationships. It is also observed that SOTA DL architectures performed better than transformer based architecture ViT. In addition, across all the evaluated metrics, VGG based architectures performed better than any other architecture used in this study.

\section{Conclusion}

The aim of this study was to collect and preprocess a new dataset of digital mammograms and radiological reports and give a baseline of SOTA DL and transformer-based architecture combined with either ANN or LSTM using a multimodal fusion approach. The highest accuracy was achieved by the VGG19  model in conjunction with ANN. For precision, the top score was again attained by VGG19 associated with ANN, followed by the combined model of ViT and LSTM. The best sensitivity score was recorded by the VGG16 with LSTM, followed by the combined models of VGG19 and LSTM fused model. The highest F1 scores were achieved by VGG19 in conjunction with LSTM, followed by the joint model of VGG19 and ANN. The best AUC score was achieved by VGG16 and LSTM combination. The VGG16 and ANN combination secured the second best AUC score. Our observations indicated that models combining LSTM with DL architectures either performed better or comparably to those using ANN in various metrics. The overall performance of the joint ViT model with LSTM or ANN was less as compared to SOTA DL architectures in most of the evaluation metrics. The results suggest that among the five evaluation metrics, SOTA DL models performed better in all evaluation metrics, while the transformer architecture along with LSTM was able to achieve second best in a precision metric only. In addition, the VGG16 and VGG19 outperformed all the SOTA DL architectures and ViT across all metrics which suggest the ability of VGG architectures in classifying the positive and negative BC instances across all the evaluation metrics. These findings further indicate that incorporating metadata extracted from radiological reports alongside images can enhance the model's performance in predicting BC. This approach has potential applications in various medical fields due to the inherently multimodal nature of healthcare data.

\begin{credits}
\subsubsection{\ackname} The authors would like to thank the Tecnológico de Monterrey and CONAHCYT for supporting their studies.

\subsubsection{\discintname}

\end{credits}
%
%
%
\bibliographystyle{splncs04}
\bibliography{ref}
%

\end{document}